\documentclass[twocolumn,amssymb,showpacs,amsmath,nobibnotes,nofootinbib,aps,superscriptaddress]{revtex4}
\usepackage{graphicx}
\usepackage{dcolumn}
\usepackage{bm}

\begin{document}
\topmargin   0mm
\preprint{APS/123-QED}
\title{Magnetic excitations in the \Kag\ staircase compounds}
\newcommand{\Kag}{Kagom\'{e}}
\newcommand\OP{O.A.~Petrenko}
\newcommand\GB{G.~Balakrishnan}
\newcommand\BF{B.~F{\aa}k}
\newcommand\NVO{$\rm Ni_3V_2O_8$}
\newcommand\CVO{$\rm Co_3V_2O_8$}
\newcommand\PRB[3]{Phys. Rev. B {#1} ({#3}), {#2}}
\newcommand\JPCM[3]{J. Phys.: Cond. Matt. {#1} ({#3}), p. {#2}}
\newcommand\SSC[3]{Solid State Comm. {#1} ({#3}), p. {#2}}
\newcommand\PRL[3]{Phys. Rev. Lett. {#1} ({#3}), {#2}}

\author{N. R. Wilson}
\email{Nicola.R.Wilson@warwick.ac.uk}
\author{O. A. Petrenko}
\author{G. Balakrishnan}
\affiliation{Department of Physics, University of Warwick, Coventry, CV4 7AL, UK}
\author{P. Manuel}
\affiliation{ISIS Facility, Rutherford Appleton Laboratory, Chilton, Didcot, OX11 0QX, UK}
\author{B. F{\aa}k}
\affiliation{CEA Grenoble, DRFMC/SPSMS, 38054, Grenoble, Cedex 9, France}

\received{6 June 2006}
\begin{abstract}
Inelastic neutron scattering measurements have been performed on single crystal samples of \CVO\ and \NVO.
The magnetic system in these compounds is believed to be frustrated, as the magnetic ions (Co$^{2+}$ with $S=3/2$ and Ni$^{2+}$ with $S=1$) adopt a buckled version of the \Kag\ lattice. 
Magnetic excitations have been observed in both samples  using a time-of-flight neutron spectrometer.
The excitation spectrum is dispersive for both samples and has a considerable gap in the low temperature phases, while the intermediate temperature phases are marked by a significant softening of the excitations energy.
\end{abstract}

\maketitle

Antiferromagnetic systems based on a \Kag\ lattice have traditionally attracted considerable interest from the theoreticians working in the field of frustrated magnetism.
The combination of a depleted triangular lattice of loosely connected corner-sharing magnetic moments and the antiferromagnetic sign of the exchange interactions gives rise to enhanced degeneracy of the ground state and to the presence of the macroscopic number of soft modes.
These factors destroy a conventional Ne\'{e}l-type magnetic order and results in a spin liquid state.
Finding real examples of the 2D \Kag\ lattice has proved to be problematic, as the residual 3D or further neighbour interactions, chemical imperfections, as well as lattice distortions and strong magnetic anisotropy tend to stabilise long range order instead of a spin liquid state.
The ground state in such situations is very sensitive to even weak interactions and often displays a complex dependence as a function of temperature and external magnetic field.

Here we report the observation of magnetic excitations in two compounds, \NVO\ and \CVO, which have recently been reported to adopt a buckled version of the \Kag\ lattice called the \Kag\ staircase \cite{Rogado}.
The lower symmetry of the staircase magnetic layers and further neighbour interactions cause a reduction in the geometrical frustration and establish long-range magnetic order in these magnets. 
An additional interest in \NVO\ is associated with the low-temperature ferroelectricity observed in this compound \cite{Lawes_PRL_2005} and the related complex behaviour caused by the magnetoelectric interactions \cite{Kenzelmann_PRB_2006}.

Large (up to 2~cm$^3$ in volume) single crystals of \CVO\ and \NVO\ were grown by the floating zone technique, using an infrared image furnace \cite{GB_JPCM_2004}.
The measurements of susceptibility and specific heat performed on these samples have shown that they have transition temperatures which agree well the previously reported values \cite{Rogado}.
Neutron scattering measurements were performed using the PRISMA time-of-flight inverse geometry crystal-analyser spectrometer at the ISIS pulsed neutron source.
The data were collected simultaneously by five double analyser-detector systems measuring five different $Q(\omega)$ lines in reciprocal space.
The \NVO\ sample was aligned with the $c$-axis vertical, defining the scattering plane as (hk0); for the \CVO\ sample the $a$-axis was kept vertical, so that the scattering plane was (0kl).
The observed intensity has been normalised using a standard vanadium calibration.
The experiments were performed in the temperature interval 1.3 to 25~K.

Figs. 1 and 2 show the maps of inelastic neutron scattering (where the intensity is colour coded) measured in the low-temperature phases of \NVO\ and \CVO\ around their respective magnetic zone centres.
For \NVO\ (see Fig. 1) the measurement is performed along the $[h30]$ direction passing through the magnetic Bragg peak at (130).
At least two excitation branches are clearly seen at $T=1.3$~K.
The lower energy branch (observed around 1~meV) is much more intense than the upper branch at about 3.5~meV.
Background measurements performed at $T=20$~K  (the ordering temperature in this compound is 9.1~K \cite{Rogado}) have confirmed that both of these branches are of magnetic origin.
Our attempt to find the spin-waves in the high temperature incommensurate phase of \NVO\ was less successful, as the scans performed at $T=7.5$~K around the $(0.73,3,0)$ magnetic Bragg peak have not revealed any excitations of significant intensity.

\begin{figure}[t]
\vspace{0.1cm}
\includegraphics[width=0.99\columnwidth]{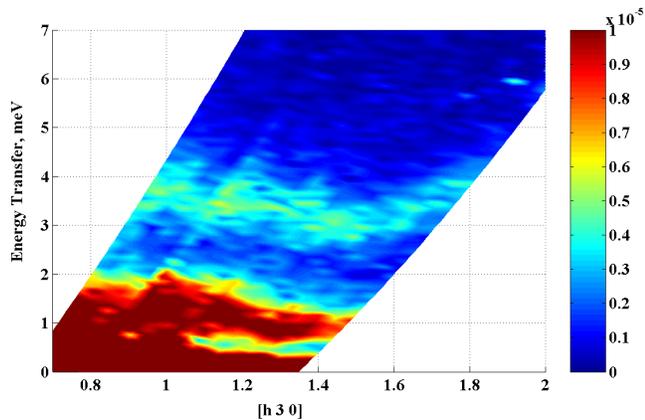}
\caption{Map of the inelastic neutron scattering intensity measured at $T=1.3 $~K in a single crystal of \NVO\ around a magnetic zone centre $(130)$.}
\label{Fig1}
\end{figure}

According to the single crystal neutron diffraction measurements \cite{Kenzelmann_PRB_2006,Lawes_PRL_2004}, the low-temperature ground state of \NVO\ is a commensurate canted antiferromagnet with the magnetic moments described by a mixture of two irreducible representations.
It would be interesting to compare the observed excitation spectrum in this state with the theoretical predictions based on the proposed model of hierarchy of magnetic forces in \NVO\ \cite{Kenzelmann_PRB_2006,Lawes_PRL_2004}, which includes nearest and next-nearest exchange, easy-axis anisotropy and Dzyaloshinskii-Moriya interactions.

\begin{figure}[t]
\vspace{0.1cm}
\includegraphics[width=0.95\columnwidth]{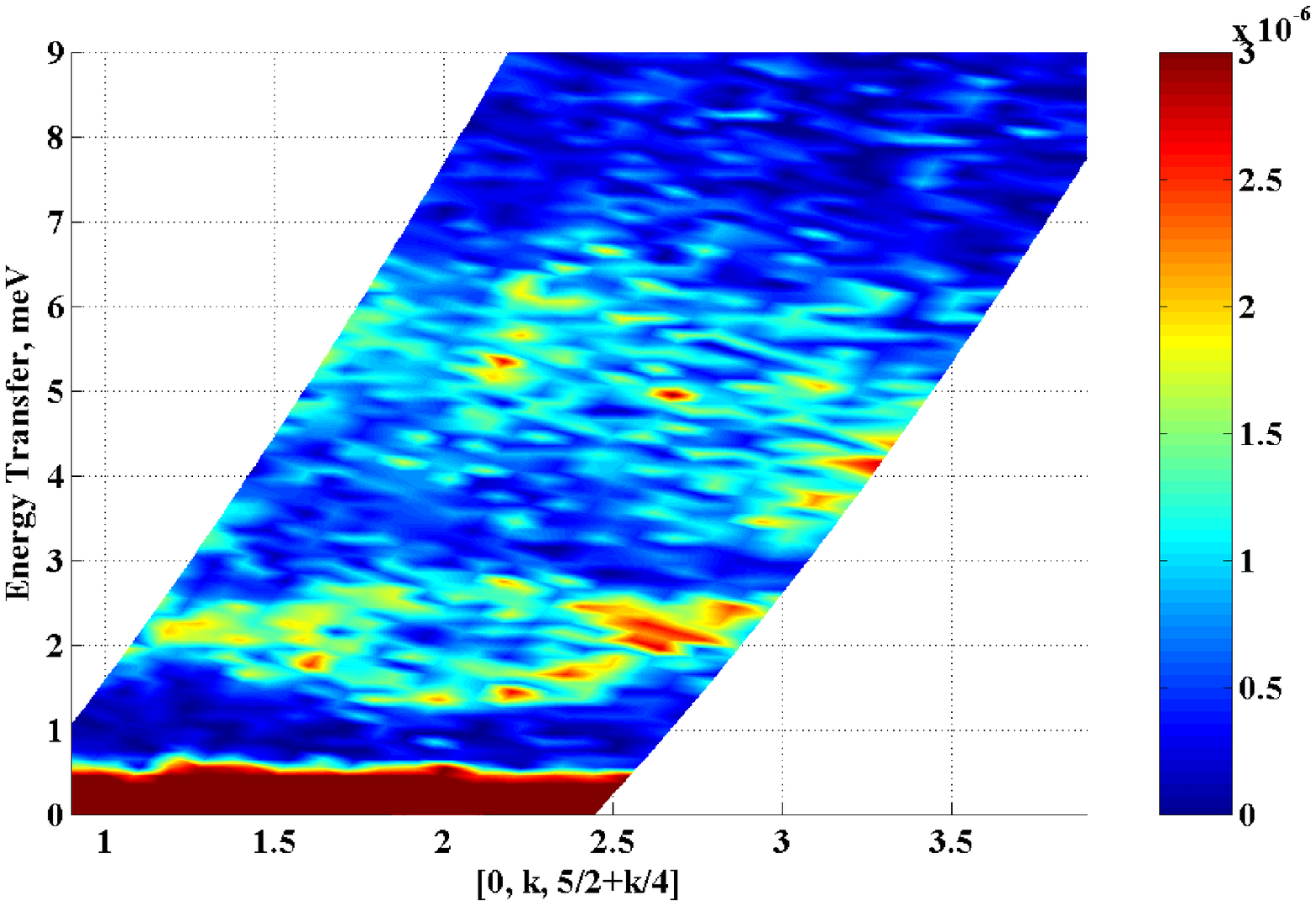}
\caption{Map of the inelastic neutron scattering intensity measured at $T=1.3 $~K in a single crystal of \CVO\ around a magnetic zone centre $(023)$.}
\label{Fig2}
\end{figure}

For \CVO,  the scans were performed either in the $[0kk]$ direction around the magnetic (040) peak or in the direction parallel to the $[0k\frac{k}{4}]$ line \cite{PRISMA_scans} and passing through the ferromagnetic Bragg peak at (023) (as shown in Fig. 2).
According to our powder neutron diffraction measurements \cite{NW_GEM_2006}, the magnetic structure of \CVO\ at temperatures below 6~K is nearly ferromagnetic with magnetic moments aligned along the $a$-axis and perhaps slightly tilted (by about 5$^\circ$) along the $c$-axis.
A peculiar feature of the magnetic ground state is the observed splitting of the two Co$^{2+}$ sites according to the value of average magnetic moment they possess.
Two thirds of the Co ions are located in the position $(\frac{1}{4},y,\frac{k}{4})$, where $y \approx 0.13$.
These ions are labelled as the ``spine" sites in Ref. \cite{Kenzelmann_PRB_2006} and carry a magnetic moment of 3.1~$\mu_B$.
The remaining Co ions reside in the position $(0,0,0)$ and are labelled  as the ``cross-tie" sites.
They carry a magnetic moment of only 1.8~$\mu_B$.
It is therefore not surprising that the excitation spectrum of \CVO\ consists of more than one branch.
In fact Fig.~2 suggests that four branches of the spin-waves may exist  in this compound.
For the lowest branch the observed gap at the zone centre is about 1.3~meV. The highest energy of this branch is 2.3~meV at the zone boundary.
The relatively small amplitude of the energy variation compared to the gap suggests that the strength of exchange interactions in this compound is comparable with the single-ion anisotropy.
This observation is consistent with the results of our earlier measurements of magnetic susceptibility in \CVO\ \cite{GB_JPCM_2004}, where the unusually large difference (up to a hundred times) in $\chi(T)$ was found for different directions of magnetic field.

At intermediate temperatures between 6~K and 12~K, \CVO\ is an incommensurate antiferromagnet with the ordering vector close to $(0\; 1/2\;0)$ \cite{NW_GEM_2006}.
Measurements of the excitation spectrum were attempted at $T=9$~K around the Bragg peaks $(0\;2.5\;1)$ and $(0\;1.5\;2)$.
Although the magnetic excitations are still visible, their energy is significantly lower when compared to the ferromagnetic phase, making it difficult to separate the inelastic and quasi-elastic signals.
Measurements on an instrument with improved energy resolution would be required for a detailed study of excitations in  the antiferromagnetic phase of \CVO.

We are grateful to L. C. Chapon for useful discussions. This work is supported by the EPSRC. NRW is grateful to the International Conference on Magnetism 2006 and to the Institute of Physics C~R~Barber Trust Fund and Condensed Matter and Materials Physics division for their financial support.


\begin{thebibliography}{00}
\bibitem{Rogado} N. Rogado {\it et al.}, \SSC{124}{229}{2002}.
\bibitem{Lawes_PRL_2005} G. Lawes  {\it et al.}, \PRL{95}{087205}{2005}.
\bibitem{Kenzelmann_PRB_2006} M.~Kenzelmann {\it et al.}, cond-mat/0510386.
\bibitem{GB_JPCM_2004} \GB\ {\it et al.}, \JPCM{16}{L347}{2004}.
\bibitem{Lawes_PRL_2004} G. Lawes {\it et al.}, \PRL{93}{247201}{2004}.
\bibitem{PRISMA_scans} It would be more natural to perform a scan along the $[0k0]$ rather than $[0k\frac{k}{4}]$ direction. This scan is however prohibited by the PRISMA instrument's geometry.
\bibitem{NW_GEM_2006} N.R.~Wilson, \OP\ and L.C.~Chapon, in preparation.

\end{thebibliography}
\end{document}